\documentclass[aps,prd,reprint,nofootinbib,longbibliography,showkeys]{revtex4-1}
\usepackage{blindtext}
\usepackage{mathtools}
\usepackage{xcolor}

\usepackage{mathrsfs}
\usepackage{braket}

\usepackage[utf8]{inputenc}
\usepackage[english,brazil]{babel}
\usepackage{hyperref}
\definecolor{mypink1}{rgb}{0.858, 0.188, 0.478}
\definecolor{mypink2}{RGB}{219, 48, 122}
\hypersetup{
    colorlinks=true,
    linkcolor=blue,
    filecolor=blue,      
    urlcolor=red,
    citecolor=blue,
}
\usepackage{color,colortbl}
\definecolor{LightCyan}{rgb}{0.88,1,1}
\usepackage{tensind}
\tensordelimiter{?}

\usepackage{graphicx}
\usepackage{hyperref}
\DeclareGraphicsExtensions{.bmp,.png,.jpg,.pdf}
\usepackage{amsmath}
\usepackage{courier}
\usepackage{verbatim}
\usepackage{amssymb}
\usepackage{amsfonts}
\usepackage{natbib}

\begin{document}
\title{Introdução à Cosmologia Quântica}
\author{Paola C. M. Delgado$^{1\dagger}$}

\affiliation{$^1$Faculty of Physics, Astronomy and Applied Computer Science,
Jagiellonian University, 30-348 Krakow, Poland}

\begin{abstract}
\textbf{Resumo}\\
Este artigo de revisão apresenta uma introdução à Cosmologia Quântica, incluindo os métodos matemáticos fundamentais para a abordagem canônica, alguns dos problemas conceituais existentes e a conexão dos modelos com possíveis observáveis.

\textbf{Abstract}\\
This review presents an introduction to Quantum Cosmology, including the mathematical methods essential to the canonical approach, some of the existing conceptual problems and the connection of the models to possible observables.
\end{abstract}

\maketitle

\section{Introdução}
A Cosmologia é o estudo científico do Universo como um todo, incluindo sua origem, dinâmica e formação de estruturas. Está intimamente ligada à Gravitação, a qual é atualmente descrita pela Teoria da Relatividade Geral de Einstein \cite{Mukhanov:991646, peacock:1999}.

A Mecânica Quântica, por sua vez, é a teoria que descreve a natureza em escalas atômicas e subatômicas, levando a uma discretização (quantização) de quantidades como energia, momento e momento angular de um sistema \cite{Sakurai:1167961,NelsonInterprMQ}. Em tais escalas, fenômenos não intuitivos para nós se manifestam, como por exemplo o caráter dual de onda-partícula de entidades físicas e o limite fundamental para a acurácia com a qual podemos prever, a partir de condições iniciais,  os valores de determinados pares de quantidades físicas (variáveis canonicamente conjugadas), o chamado princípio da incerteza de Heisenberg \cite{eisberg1967fundamentals}. No limite macroscópico, a Mecânica Quântica recupera os resultados da Mecânica Clássica, o que é descrito pelo princípio da correspondência e corroborado pelo teorema de Ehrenfest. 
Dessa forma, a quantização de qualquer sistema físico, inclusive em escalas macroscópicas, deveria ser viável.\footnote{Vale mencionar que, apesar de matematicamente bem estabelecida e experimentalmente verificada, a Mecânica Quântica não possui uma interpretação universalmente aceita. Existem diversas abordagens sobre de que forma a teoria se conecta à realidade que percebemos \cite{NelsonInterprMQ}. Dentre as muitas escolas de pensamento estão a interpretação de Copenhagen \cite{doi:10.1080/00107514.2015.1111978}, a interpretação de De Broglie-Bohm \cite{universe7050134,holland1995quantum,Siqueira-Batista:2022fzt}, a interpretação de muitos mundos \cite{quantum4030018} e a interpretação de histórias consistentes \cite{PhysRevA.54.2759}. Importantes questões são investigadas em cada uma dessas (e outras) abordagens, incluindo o caráter determinista ou probabilístico da teoria e o chamado problema da medida \cite{measurproblem}.}

A Cosmologia Quântica parte desse princípio de aplicabilidade da Mecânica Quântica a todos os sistemas físicos existentes, incluindo o próprio Universo. Diversos sistemas físicos compostos por campos de matéria já foram quantizados com êxito, resultando em teorias tais como a Eletrodinâmica e a Cromodinâmica Quânticas \cite{Peskin:257493,QCD}, as quais tiveram predições corroboradas por evidências experimentais \cite{Peskin:257493,PhysRevLett.97.030801,PhysRevLett.99.039902,Bethke:2006ac}. Por outro lado, a quantização do Universo requer não somente a quantização da matéria, mas também do próprio espaço-tempo. Esta é uma implicação direta da Relatividade Geral, uma vez que a teoria descreve o espaço-tempo como uma entidade física, a qual satisfaz leis dinâmicas e interage com a matéria. Por essa razão, a Cosmologia Quântica está intimamente ligada à Gravitação Quântica, a qual almeja quantizar a gravidade. Contudo, as bases matemáticas e conceituais para a quantização do espaço-tempo não são bem estabelecidas e numerosos desafios surgem em diferentes abordagens. 

Além do princípio mencionado acima, referente a quantizar o Universo enquanto um sistema físico, uma outra grande motivação para a busca de uma Teoria Quântica da Gravitação vem da própria Cosmologia clássica: a singularidade inicial, conhecida popularmente como {\it Big Bang}, evidencia a limitação da teoria em descrever regimes em energias extremamente altas. Matematicamente tal singularidade é descrita pela divergência (infinitude) da densidade de energia e da curvatura do espaço-tempo, o que as torna não físicas. Tendo em vista que singularidades geralmente apontam uma incompletude da teoria, tendo sido algumas delas previamente resolvidas através da quantização \cite{planckpaper, Peskin:257493}, é natural considerar a ideia de que uma Teoria da Gravitação Quântica possa resolver a singularidade inicial no Universo. Como será mostrado na Seção \ref{CosmologiaQuantica}, essa é de fato uma consequência de diferentes propostas para a Cosmologia Quântica \cite{Delgado:2020htr,Frion:2020tqq,CarlaOlesyaJuliobounce,Vicente:2023hba,Pinto-Neto:2021gcl,EdwardWilson-Ewing_2013,PhysRevD.97.083517}.

Ainda que uma única e bem estabelecida forma de quantizar o Universo não tenha sido encontrada até o momento, a busca por teorias nessa direção tem levado a um considerável desenvolvimento conceitual e matemático da nossa descrição sobre a gravidade e sobre a Mecânica Quântica. Dentre as numerosas propostas estão a Gravitação Quântica Canônica, incluindo a equação de Wheeler-DeWitt \cite{PhysRev.160.1113,WDWKiefer} e a Gravitação Quântica em Laços (ou em {\it Loop}) \cite{Banerjee:2011qu, EdwardWilson-Ewing_2013,LQCAgullo}, a Teoria das Cordas \cite{Mukhi_2011}, a Gravidade Assintoticamente Segura \cite{ASGAstrid} e a Triangulação Dinâmica Causal \cite{Loll_2020}. Esta não é, contudo, uma lista exaustiva das teorias existentes e mais exemplos podem ser encontrados em \cite{euclidqg,universe4100103,Oliveira-Neto:2012sjq,Oliveira-Neto:2017yui,Giacchini:2021osq,Surya:2019ndm,Henz:2013oxa,Moffat:2000gr,Taylor:1983su,Penrose:1972ia}.

Dadas as diferentes abordagens exploradas como possíveis caminhos para uma Teoria Quântica da Gravitação, diferentes cenários no contexto cosmológico podem emergir, tornando o Universo primordial um excelente regime em que tais teorias podem ser eventualmente testadas. Entretanto, a conexão com observáveis tem se mostrado um grande desafio imposto pelas escalas de energia que conseguimos acessar. Atualmente, a informação mais longínqua que temos dos primórdios do Universo é a chamada Radiação Cósmica de Fundo (geralmente referenciada como CMB, sigla da expressão em inglês {\it Cosmic Microwave Background})\footnote{A Radiação Cósmica de Fundo (CMB) é uma radiação na faixa de frequência do micro-ondas que detectamos em todas as direções. Ela se originou durante a combinação de elétrons e prótons no Universo primordial (fenômeno conhecido por recombinação na literatura), aumentando o livre caminho médio dos fótons, que passaram a viajar até nós. Antes da combinação os fótons sofriam sucessivos espalhamentos e ficavam retidos no plasma primordial, tornando o Universo opaco para nossas observações.} \cite{Durrer:2008eom} e sua escala de energia é menor (ou seja, aconteceu temporalmente depois) que a escala de energia de uma possível Gravitação Quântica. Por essa razão, é extremamente desafiador encontrar possíveis evidências observacionais dessas teorias. Apesar disso, algumas relações com observáveis já foram obtidas e vínculos já foram impostos \cite{Agullo_2021,Delgado:2021mxu,vanTent:2022vgy,K:2023gsi,Calcagni:2012vw, Cai:2014bea,Cai:2014xxa,Battefeld:2014uga}, levando inclusive à exclusão de algumas classes de modelos, como será explicado na Seção \ref{VinculosObs}. Uma outra perspectiva para testar teorias de Gravitação Quântica, antes mesmo da CMB, está na possível detecção futura de ondas gravitacionais primordiais \cite{Calcagni_2021}.  Uma vez que estas interagem muito fracamente com a matéria, elas já estariam se propagando até nós enquanto os fótons da CMB estavam retidos. Dessa forma, escalas de energia maiores poderiam ser exploradas.

As próximas seções encontram-se organizadas da seguinte maneira: a Seção \ref{QuantizaçãoRG} apresenta a abordagem canônica para a quantização da Teoria da Relatividade Geral, a qual se baseia no formalismo ADM. A Seção \ref{CosmologiaQuantica} aborda as consequências da quantização para cenários cosmológicos concretos, considerando como exemplos a Cosmologia Quântica de De Broglie-Bohm e a Cosmologia Quântica em Laços. Por fim, as relações com observáveis são exploradas na Seção \ref{VinculosObs}.

\section{Quantização da Relatividade Geral}\label{QuantizaçãoRG}
Nesta seção será apresentada a proposta de quantização canônica para uma teoria da Gravitação Quântica. Tal abordagem faz uso da Mecânica Hamiltoniana, a qual descreve sistemas físicos em termos de suas variáveis de posição e momento, e da segunda quantização, onde campos clássicos são promovidos a operadores quânticos. Na Subseção \ref{FormADM}, o formalismo Hamiltoniano da Relatividade Geral é introduzido, seguido pelo procedimento de quantização canônica que leva à equação de Wheeler-DeWitt (Subseção \ref{WDWeq}) e à base da Gravitação Quântica em Laços (Subseção \ref{LQG}). 

Daqui em diante, as coordenadas do espaço-tempo serão denotadas por $x^{\mu}$, onde $\mu=0$ se refere à coordenada temporal e $\mu=1,2,3$ às coordenadas espaciais. Índices em letras gregas tomam valores de $0$ a $3$, enquanto índices em latim vão de $1$ a $3$. Derivadas em relação às coordenadas são denotadas por $\partial_\mu\equiv \partial/\partial x^\mu$. O espaço-tempo, o qual é descrito matematicamente por uma variedade, será representado por seu tensor métrico $g_{\mu\nu}$, que nada mais é que uma estrutura nessa variedade que permite a definição de distâncias e ângulos. A assinatura da métrica usada é $(-,+,+,+)$. A notação de Einstein é utilizada nas expressões tensoriais, indicando que índices repetidos devem ser somados (por exemplo, $V^i V_i=\Sigma_{a=1}^3 V^i V_i$). Por fim, denotaremos a delta de Kronecker como $\delta_{ij}$ e o tensor de Levi-Civita como $\epsilon_{ijk}$. 

\subsection{Formulação hamiltoniana da Relatividade Geral}\label{FormADM}
A formulação hamiltoniana da Relatividade Geral foi desenvolvida por Richard Arnowitt, Stanley Deser e Charles Misner, tendo ficado conhecida como formalismo ADM \cite{PhysRev.116.1322}. A formulação se baseia em quantidades geométricas que caracterizam uma foliação de hipersuperfícies espaciais\footnote{Hipersuperfícies são variedades algébricas de dimensão $n-1$ inseridas em um espaço de dimensão $n$. No caso da Relatividade Geral, as variedades englobam as $3$ dimensões espaciais e o tempo é descrito como a dimensão extra no espaço com $n=4$.} na direção do tempo. 

\begin{figure}[h]
\centering
\includegraphics[width=.9\columnwidth]{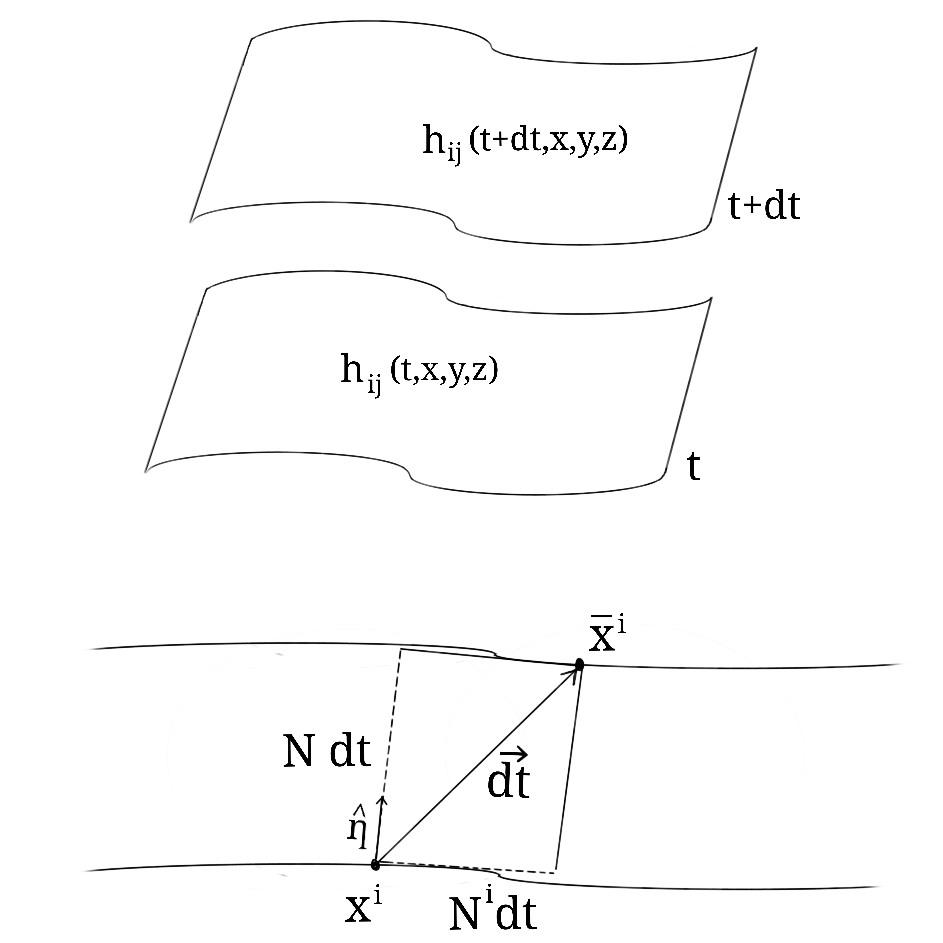}
\caption{Duas hipersuperfícies separadas na direção temporal por um intervalo infinitesimal $dt$. A função lapso $N$ e o vetor deslocamento $N^i$ são representados geometricamente. A normal $\eta_\mu$ à superfície $h_{ij}(t,x,y,z)$ é representada através de seu correspondente unitário $\hat{\eta}$.}
\label{figadm}
\end{figure}

Tais hipersuperfícies são definidas através da constância de uma função $f$ das coordenadas, ou seja, $f(x^{\mu})={\rm constante}$ e suas normais $\eta_\mu$. Ao introduzir a coordenada temporal $t\equiv x^0$, podemos escrever $\eta_\mu=-N\delta_\mu^0$, onde N é chamada de função lapso e é normalizada através de $g^{\mu\nu}\eta_\mu \eta_\nu=-1$. Podemos também definir o projetor $h^{\mu\nu}\equiv g^{\mu\nu}+\eta^\mu \eta^\nu$, de forma que sua matriz inversa $h_{\mu\nu}$ é o tensor métrico das hipersuperfícies. Por fim, definimos o chamado vetor deslocamento $N^{i}\equiv g^{i0}N^2$, o qual descreve a taxa de mudança do deslocamento de $x^i$ de uma hipersuperfície para outra. Na Figura \ref{figadm} tais definições são apresentadas de forma geométrica. A função lapso, o vetor deslocamento e a métrica das hipersuprefícies são então utilizadas para descrever a métrica do espaço-tempo quadridimensional em questão:
\begin{equation}
 g^{\mu\nu}=\begin{pmatrix}
-\frac{1}{N^2} & \frac{N^i}{N^2} \\
 \frac{N^j}{N^2}& h^{ij}-\frac{N^i N^j}{N^2} \\
\end{pmatrix}.
\end{equation}

Uma importante quantidade no contexto da geometria diferencial ao se considerar hipersuperfícies inseridas em uma variedade é a curvatura extrínseca. No presente cenário considerado, tal quantidade é dada por 
\begin{equation}
    K_{ij}=\frac{1}{2N}\left[\dot{h}_{ij}-\nabla_i N_j-\nabla_j N_i\right],
\end{equation}
onde $\nabla_\mu V_\nu \equiv \partial_\mu V_\nu - \Gamma^\lambda_{\mu\nu} V_\lambda$ é a derivada covariante do espaço-tempo em questão (derivada ao longo de vetores tangentes da variedade) e $\nabla_i$ é a derivada covariante tridimensional. A quantidade $\Gamma^\lambda_{\mu\nu}\equiv \frac{1}{2}g^{\lambda \rho}\left(\partial_\nu g_{\rho\mu}+\partial_\mu g_{\rho\nu}-\partial_\rho g_{\mu\nu}\right)$ representa os símbolos de Christoffel, os quais descrevem a conexão afim do espaço-tempo. O ponto representa a derivada temporal.

A densidade lagrangiana desse sistema pode ser escrita em termos das quantidades anteriormente definidas
\begin{equation}\label{LADMeq}
    \mathcal{L}=N h^{\frac{1}{2}}\left(R^{(3)}+K^{ij}K_{ij}-K^2\right),
\end{equation}
sendo $R^{(3)}$ o escalar de Ricci das hipersuperfícies, o qual quantifica a curvatura da variedade, $K\equiv K_i^i$ e $h$ o determinante da métrica das hipersuperfícies. Uma vez que essa densidade lagrangiana não depende de $\partial_0 N$ ou $\partial_0 N^i$, os momentos conjugados do lapso $N$ e da função deslocamento $N^i$ são nulos. Tais variáveis não dinâmicas devem ser incluídas na ação do sistema como fatores multiplicativos dos vínculos, os chamados multiplicadores de Lagrange. Dessa forma, a ação gravitacional $S$ pode ser escrita como
\begin{eqnarray}
\nonumber    S&=&\frac{1}{16\pi}\int\left[\Pi^{ij}\dot{h}_{ij}+N_i 2 \nabla_i \Pi^{ij}\right.+\\
\nonumber    &-&\left.N\left(G_{ijkl}\Pi^{ij}\Pi^{kl}-h^{\frac{1}{2}}R^{(3)}\right)\right] dt d^3x,\\
\nonumber    \Pi^{ij}&\equiv& \frac{\delta \mathcal{L}}{\delta(\partial_0 h^{ij})}=-h^{\frac{1}{2}}\left(K_{ij}-h_{ij} K\right),\\
    G_{ijkl}&\equiv&\frac{h^{-\frac{1}{2}}}{2}\left(h_{ik}h_{jl}+h_{il}h_{jk}-h_{ij}h_{kl}\right). \label{GdeWitt}
\end{eqnarray}
As quantidades acompanhadas dos multiplicadores de Langrange $N$ e $N_i$ são os vínculos secundários da teoria e são respectivamente chamados de super-hamiltoniana $\mathcal{H}$ e supermomento $\mathcal{H}^j$\footnote{A igualdade fraca denotada por $\approx$ se deve ao fato das equações serem satisfeitas somente quando os vínculos são aplicados.}
\begin{eqnarray}
    \mathcal{H}&\equiv& G_{ijkl}\Pi^{ij}\Pi^{kl}-h^{\frac{1}{2}}R^{(3)}\approx 0,\\
    \mathcal{H}^j&\equiv& -2 \nabla_i \Pi^{ij}\approx 0.
\end{eqnarray}
O primeiro está relacionado à covariância da teoria sob transformações gerais do tempo, enquanto o segundo descreve a covariância sob transformações de coordenadas espaciais.  

Por sua vez, a densidade hamiltoniana $\mathscr{H}\equiv\Pi^{ij} \dot{h}_{ij}-\mathcal{L}$ toma a seguinte forma
\begin{equation}\label{denHam}
    \mathscr{H}=\int \left(N \mathcal{H}+N_i \mathcal{H}^i\right) dt d^3x.
\end{equation}

\subsection{Quantização canônica e a equação de Wheeler–DeWitt}\label{WDWeq}
Como vimos na seção anterior, a Teoria da Relatividade Geral é covariante sob transformações de coordenadas. Em outras palavras, as leis físicas tomam a mesma forma em todos os sistemas referenciais. Matematicamente, essa propriedade leva aos vínculos, os quais reduzem os graus de liberdade do sistema. A quantização de sistemas vinculados foi desenvolvida por Paul Dirac \cite{Dirac:2001:LQM} e proporcionou a base para o procedimento da quantização canônica da Relatividade Geral. 

Primeiramente promovemos as variáveis canônicas da teoria a operadores quânticos, de forma que os parênteses de Poisson $\{X,Y\}$ serão identificados como comutadores, ou seja $i\hbar\{X,Y\}\equiv [\hat{X},\hat{Y}]$. Dessa forma, a métrica das hipersuperfícies $h_{ij}$ se torna um operador $\hat{h}_{ij}$, o qual atua em funcionais de onda $\Psi$.

A partir da densidade hamiltoniana \eqref{denHam} podemos escrever a seguinte equação funcional de Schr{\"o}dinger\footnote{A equação de Schr{\"o}dinger $\hat{H}\ket{\Psi(t)}=i\hbar\frac{d}{dt}\ket{\Psi(t)}$ é a equação que governa a dinâmica da função de onda $\Psi(t)$ em um sistema quântico.}
\begin{equation}\label{Seq}
    i \partial_0 \Psi=\int \left(N\hat{\mathcal{H}}+N_i \hat{\mathcal{H}^i}\right)\Psi d^3x,
\end{equation}
onde o lado direito da equação representa a hamiltoniana $H$. Já os vínculos da super-hamiltoniana e do supermomento tomam a seguinte forma
\begin{eqnarray}
    \hat{\mathcal{H}}\Psi &=& 0,\label{Ceq1}\\
    \hat{\mathcal{H}^i}\Psi &=& 0. \label{Ceq2}
\end{eqnarray}
O funcional de onda deve, então, satisfazer não somente a equação \eqref{Seq}, mas também os vínculos \eqref{Ceq1} e \eqref{Ceq2}. A equação \eqref{Ceq1} é a chamada equação de Wheeler-DeWitt, enquanto \eqref{Ceq2} é o chamado vínculo de difeomorfismo. 

\subsection{Aspectos introdutórios da Gravitação Quântica em Laços}\label{LQG}
A Gravitação Quântica em Laços realiza a quantização também de forma canônica, mas fazendo uso de quantidades inspiradas em teorias de {\it gauge} \cite{Banerjee:2011qu, EdwardWilson-Ewing_2013,LQCAgullo}. Por essa razão, o espaço de fase passa a ser descrito por uma conexão de {\it gauge} do grupo SU(2)\footnote{Na chamada teoria de grupos, um grupo é uma estrutura constituída por um conjunto de elementos e uma operação que satisfazem as propriedades de fechamento, associatividade, identidade e elemento inverso. O grupo SU(2), onde SU faz referência a {\it Special Unitary}, é o grupo das matrizes $2\times2$, unitárias, complexas e com determinante igual a $1$, estando relacionado a rotações no espaço tridimensional.} $A^{\bar{i}}_i$ e pelo seu momento canonicamente conjugado $E^{\bar{i}}_i$. Os índices com barras são índices do grupo SU(2), os quais descrevem os graus de liberdade extras que surgem nessa formulação. Tais variáveis são escritas em termos da chamada co-tríade $e^{\bar{i}}_i$, a qual se relaciona à métrica $h_{ij}$ através de
\begin{eqnarray}
    h_{ij}\equiv e^{\bar{i}}_i e^{\bar{j}}_j \delta_{\bar{i}\bar{j}}.
\end{eqnarray}
Dessa forma, escrevemos
\begin{eqnarray}
    A^{\bar{i}}_{i}&=&\Gamma^{\bar{i}}_{i}+\gamma K^{\bar{i}}_{i},\\
    E^{i}_{\bar{i}}&=&\sqrt{h}e^{i}_{\bar{i}},
\end{eqnarray}
onde $\gamma$ é um parâmetro real, $h$ é o determinante de $h_{ij}$, $K^{\bar{i}}_{i}$ se relaciona à curvatura extrínseca através de $K^{\bar{i}}_{i}=K_{ij} e^{j}_{\bar{j}}\delta^{\bar{i}\bar{j}}$ e $\Gamma^{\bar{i}}_{i}$ é a chamada conexão de {\it spin} dada por $\nabla_j E^{i}_{\bar{i}}+\epsilon_{\bar{i}\bar{j}\bar{k}}\Gamma^{\bar{j}}_j E^{i\bar{k}}=0$.

Como mostrado na Seção \ref{FormADM}, desejamos obter os vínculos da teoria, os quais estão relacionados às simetrias de {\it gauge}. Similarmente à abordagem anterior, temos o vínculo da super-hamiltoniana e o vínculo do supermomento. Entretanto, note que os graus de liberdade extras que introduzimos na presente formulação levam a uma outra liberdade de {\it gauge}, a qual está relacionada à invariância de $\delta_{\bar{i}\bar{j}}$ frente a rotações SU(2). Por essa razão, encontramos um vínculo extra, conhecido como vínculo de Gauss. É possível mostrar que os três vínculos mencionados são dados por
\begin{eqnarray}\label{vincLQG}
 \nonumber   \mathcal{H}&=&\frac{1}{\sqrt{\bar{E}}}\epsilon_{\bar{i}\bar{j}\bar{k}}\left[F^{\bar{i}}_{ij}-(1+\gamma^2)\epsilon^{\bar{i}}_{\bar{l}\bar{m}}K^{\bar{l}}_i K^{\bar{m}}_j\right]\times\\
\nonumber &\times& E^{i\bar{j}} E^{j\bar{k}},\\
\nonumber    \mathcal{H}_i&=&F^{\bar{i}}_{ij}E^j_{\bar{i}},\\
    \mathcal{G}_{\bar{i}}&=&\partial_i E_{\bar{i}}^i+\epsilon_{\bar{i}\bar{j}\bar{k}}\Gamma^{\bar{j}}_i E^{i\bar{k}},
\end{eqnarray}
onde $\bar{E}\equiv |det(E)|$ e $F^{\bar{i}}_{ij}=\partial_i A^{\bar{i}}_j-\partial_j A^{\bar{i}}_i+\epsilon^{\bar{i}}_{\bar{j}\bar{k}}A^{\bar{j}}_i A^{\bar{k}}_j$.

Seguindo o procedimento de quantização canônica, desejamos promover variáveis clássicas a operadores quânticos. Para isso, são utilizadas as holonomias\footnote{Uma holonomia é um objeto matemático que descreve a variação de um tensor quando este é transportado ao longo de uma curva em um espaço-tempo curvo.} de $A^{\bar{i}}_i$ e o fluxo de $E^i_{\bar{i}}$. Tais quantidades são independentes da métrica do espaço-tempo e invariantes frente a difeomorfismos. Além disso, são definidas as redes de {\it spin}, as quais conferem ao espaço-tempo uma estrutura discreta. Tais redes de {\it spin} estão relacionadas ao chamados {\it loops}, os quais se referem à unidade fundamental do espaço-tempo.
Os detalhes sobre esse procedimento de quantização podem ser encontrados em \cite{Banerjee:2011qu}.

\section{Quantização do Universo}\label{CosmologiaQuantica}
Nessa seção iremos explorar como a proposta canônica para uma teoria da Gravitação Quântica é aplicada no contexto cosmológico, o qual goza de simetrias a serem satisfeitas, levando ao que conhecemos como Cosmologia Quântica. 

\subsection{Minissuperespaço preenchido por um fluido perfeito}
Superespaço é o nome dado ao espaço das hipersuperfícies tridimensionais $h_{ij}$, o qual possui dimensão infinita. Por sua vez, um minissuperespaço é uma forma reduzida do superespaço, obtido através da redução dos graus de liberdade do sistema através do uso de simetrias. Tais simetrias são advindas da homogeneidade e isotropia do Universo. 

A fim de incorporar as simetrias do sistema, consideramos a função lapso homogênea, ou seja, $N=N(t)$ e o vetor deslocamento $N^i=0$, indicando a isotropia do espaço-tempo\footnote{Tais condições para a função lapso e o vetor deslocamento são usadas no chamado nível de fundo, quando perturbações cosmológicas ainda não são consideradas.}. Dessa forma, o elemento de linha\footnote{O elemento de linha pode ser entendido como o segmento de linha associado a um vetor de deslocamento infinitesimal em um espaço métrico. Seus termos estão diretamente relacionados ao tensor métrico através de $ds^2=g_{\mu\nu}dx^{\mu}dx^{\nu}$.} pode ser escrito como
\begin{equation}
    ds^2=-N^2(t) dt+h_{ij}(x,t) dx^i dx^j.
\end{equation}
A métrica $h_{ij}$, por sua vez, pode ser restrita a 
\begin{equation}
    h_{ij}(x,t) dx^i dx^j=a^2(t)d\Omega_3^2,
\end{equation}
onde $d\Omega_3^2$ é o elemento de linha de uma triesfera e $a(t)$ é o chamado fator de escala, o qual parametriza a expansão do Universo. De forma mais genérica, é possível restringir $h_{ij}$ com uma quantidade finita de parâmetros $q^\alpha(t)$, com $\alpha=1,...,n$. No presente caso, $q^1(t)=a(t)$ e os demais $q^\alpha$ correspondem aos graus de liberdade de matéria. Os momentos conjugados  aos parâmetros $q^\alpha$ serão denotados por $p_\alpha$.

Assim as quantidades utilizadas na formulação hamiltoniana da Relatividade Geral podem ser escritas em termos de $N(t)$ e $a(t)$. A ação da teoria, incluindo a densidade lagrangiana de matéria $\mathcal{L}_{\rm M}$ em termos de campos $\phi^A$, toma então a forma
\begin{eqnarray}
\nonumber S&=&\int N h^{\frac{1}{2}}\left(R^{(3)}+K^{ij}K_{ij}-K^2\right)dt d^3x+\\
\nonumber &+&\int \mathcal{L}_{\rm M}\left(\phi^A,h_{ij},N_i,N\right)N h^{\frac{1}{2}} dt d^3x\\
    &=&\int_0^1 dt N\left[\frac{1}{2N^2} f_{\alpha\beta}(q)\dot{q}^\alpha \dot{q}^\beta-U(q)\right],\label{mini}
\end{eqnarray}
onde $f_{\alpha\beta}(q)$ é $G_{ijkl}$ definido em \eqref{GdeWitt} reduzido ao minissuperespaço e $V(q)$ é uma função de $q$ que pode ser entendida como um potencial. Os limites de integração $0$ e $1$ são obtidos ao ajustar a função lapso e o tempo apropriadamente. Note que a ação \eqref{mini} corresponde à descrição de uma partícula relativística em um espaço-tempo curvo, contendo as contribuições cinética e potencial. Dessa forma, a difícil tarefa de resolver a equação de Wheeler-DeWitt \eqref{Ceq1} e o vínculo de difeomorfismo \eqref{Ceq2} no superespaço é simplificada ao problema de uma partícula no minissuperespaço. Por uma questão de consistência, as equações de movimento que podem ser obtidas dessa ação devem corresponder às equações de Einstein, as quais descrevem a Teoria da Relatividade Geral no regime clássico. 

A hamiltoniana $H=p_\alpha \dot{q}^\alpha-L$ correspondente é dada por
\begin{equation}
    H = N\left[\frac{1}{2}f^{\alpha\beta} p_\alpha p_\beta+U(q)\right],
\end{equation}
de onde obtemos o vínculo da super-hamiltoniana
\begin{equation}
    \frac{1}{2}f^{\alpha\beta} p_\alpha p_\beta+U(q)\approx 0.
\end{equation} 

Tradicionalmente os modelos de minissuperespaço têm sido entendidos como uma aproximação para extrair informações do Universo como um todo. Entretanto, não existe uma confirmação de que essa abordagem leve a uma representação fidedigna e completa da teoria. Uma alternativa é interpretar a quantização do minissuperespaço como a quantização da menor unidade representativa do espaço-tempo, abordagem conhecida como {\it single-patch}\cite{Bojowald_2015}. Neste artigo adotaremos a abordagem mais tradicional, na qual o Universo é representado pelo minissuperespaço.

Considerando o elemento de linha de um Universo homogêneo e isotrópico, também conhecido como elemento de linha de Friedmann-Lemaître-Robertson-Walker (FLRW) (em coordenadas esféricas $r,\theta,\Phi$ e com curvatura espacial $k=-1,0,+1$)
\begin{eqnarray}
  \nonumber  ds^2&=&-N^2 dt^2+a^2\left(\frac{dr^2}{1-kr^2}+r^2 d\theta^2+\right.\\
    &+&\left.r^2\sin^2\theta d\Phi^2\right),
\end{eqnarray}
obtemos a densidade  lagrangiana \eqref{LADMeq} no formalismo ADM 
\begin{eqnarray}
    \mathcal{L}&=&\frac{\ddot{a}a^2}{N}-\frac{\dot{a}\dot{N}a^2}{N^2}+\frac{6\dot{a}^2a}{N}+kNa,\\
    &=&-\frac{a\ddot{a}^2}{N}+kaN
\end{eqnarray}
tendo sido utilizada integração por partes na última igualdade. 

A hamiltoniana do sistema, por sua vez, é dada por
\begin{equation}\label{Hgrav}
    H=N\left(-\frac{P_a^2}{4a}-6ka\right),
\end{equation}
onde $P_a=-2a\dot{a}/N$ é o momento canonicamente conjugado ao fator de escala $a$.

Consideremos agora que o minissuperespaço está preenchido por um fluido perfeito descrito pela seguinte lagrangiana de matéria
\begin{equation}
    L_M=\sqrt{-g}\left(\frac{1}{2}g^{\mu\nu}\partial_\mu \phi \partial_\nu \phi\right)^n,
\end{equation}
onde $n$ é um número inteiro e $\phi$ é um campo escalar relacionado à quadrivelocidade do fluido 
\begin{equation}
    U_\mu=\frac{\partial_\mu\phi}{\sqrt{g^{\mu\nu}\partial_\mu\phi \partial_\nu \phi}}.
\end{equation}

O parâmetro da equação de estado do fluido pode ser obtido através da definição do tensor momento-energia\footnote{Para um fluido perfeito, o tensor momento-energia pode ser escrito como $T_{\mu\nu}=(\rho+P)U_\mu U_\nu-P g_{\mu\nu}$, onde $\rho$ é a densidade de energia e $P$ é a pressão. Por outro lado, o parâmetro da equação de estado é dado por $\omega=P/\rho$. Combinando tais expressões com a definição \eqref{tmunu}, obtemos $\omega$ em função de $n$.} 
\begin{equation}\label{tmunu}
    T_{\mu\nu}=\frac{2}{\sqrt{-g}}\frac{\partial L_M}{\partial g^{\mu\nu}}
\end{equation}
e é dado por
\begin{equation}
    \omega=\frac{1}{2n-1}.
\end{equation}

A hamiltoniana do sistema é então escrita como
\begin{equation}
    H_M=\frac{1}{\omega (\sqrt{2}n)^{1+\omega}}N \frac{p_\phi^{1+\omega}}{a^{3\omega}},
\end{equation}
onde $p_\phi$ representa o momento canonicamente conjugado a $\phi$. Ao considerar a seguinte transformação de coordenadas
\begin{eqnarray}
    T = \frac{\omega (\sqrt{2}n)^{1+\omega}}{1+\omega}\frac{\phi}{p_\phi^{1+\omega}},\\
    P_T = \frac{1}{\omega} \left(\frac{p_\phi}{\sqrt{2}n}\right)^{1+\omega},
\end{eqnarray}
obtemos 
\begin{equation}\label{Hmatter}
    H_M=N\frac{P_T}{a^{3\omega}}.
\end{equation}
Uma derivação alternativa desse resultado é obtida em \cite{PhysRevD.2.2762}.

Considerando as hamiltonianas obtidas para a parte gravitacional \eqref{Hgrav} e para a parte de matéria \eqref{Hmatter}, chegamos na descrição de um minissuperespaço homogêneo e isotrópico, com curvatura espacial $k=0$ e preenchido por um fluido perfeito:
\begin{equation}\label{HTOTAL}
    H=N\left(-\frac{P_a^2}{4a}+\frac{P_T}{a^{3\omega}}\right).
\end{equation}

\subsection{Cosmologia Quântica de De Broglie-Bohm}\label{cosmdbb}
Nesta seção trataremos da quantização do Universo de acordo com a interpretação de De Broglie-Bohm, a qual possui um caráter determinístico, sendo as propriedades probabilísticas da Mecânica Quântica meramente estatísticas. Tal interpretação faz uso das chamadas variáveis ocultas, as quais determinam o desenvolvimento do sistema quântico. As previsões da teoria concordam com a tradicional interpretação de Copenhagen, desde que a equação de Schr{ö}dinger seja mantida. Nesta seção serão introduzidas brevemente as quantidades necessárias para realizar a quantização. Uma abordagem mais completa da interpretação pode ser encontrada em \cite{universe7050134}.

Primeiramente aplicaremos a quantização de Dirac à hamiltoniana \eqref{HTOTAL} ao promovê-la a um operador quântico e ao fazer o uso da equação \eqref{Ceq1}. Ao optar por um ordenamento de operadores\footnote{A necessidade de escolher um ordenamento está relacionada ao princípio de incerteza de Heisenberg, de acordo com o qual dois operadores complementares não podem ser medidos simultaneamente. Matematicamente tal fato é descrito pela não comutatividade de tais operadores. Na equação \eqref{WDWcos0} o fator de escala $a$ e seu momento conjugado $\partial/\partial a$ não comutam, tornando a escolha do ordenamento necessária.}, obtemos
\begin{equation}\label{WDWcos0}
    i\frac{\partial}{\partial T}\Psi=\frac{a^{\frac{3\omega-1}{2}}}{4} \frac{\partial}{\partial a}\left[a^{\frac{3\omega-1}{2}}\frac{\partial}{\partial a}\right]\Psi,
\end{equation}
a qual representa a equação de Wheeler-DeWitt. Como veremos na Seção \eqref{ProbOfTime}, a variável $T$, relacionada ao fluido perfeito, pode ser entendida como o tempo. 

Na interpretação de De Broglie-Bohm, escrevemos a função de onda como $\Psi=R \exp[iS]$, sendo $R$ a amplitude e $S$ a fase da onda. Definindo $\rho\equiv a^{\frac{3\omega-1}{2}} |\Psi|^2$, obtemos que \eqref{WDWcos0} resulta em duas equações:
\begin{eqnarray}
  \nonumber  &&\frac{\partial S}{\partial T}-\frac{a^{3\omega-1}}{4}\left(\frac{\partial S}{\partial a}\right)^2+\\
    &+&\frac{a^{\frac{3\omega-1}{2}}}{4R}\frac{\partial}{\partial a}\left[a^{\frac{3\omega-1}{2}}\frac{\partial R}{\partial a}\right]=0,\label{eqbohm1}\\
    &&\frac{\partial \rho}{\partial T}-\frac{\partial}{\partial a}\left[\frac{a^{3\omega-1}}{2}\frac{\partial S}{\partial a}\rho\right]=0\label{eqbohm2}.
\end{eqnarray}
Além disso, o espaço de configurações possui um caráter determinístico, sendo descrito pela chamada equação guia
\begin{equation}\label{eqguia}
    \dot{a}=-\frac{a^{3\omega-1}}{2}\frac{\partial S}{\partial a}.
\end{equation}
Note que a equação \eqref{eqbohm1} toma a forma de uma equação de Hamilton-Jacobi no âmbito da Mecânica Quântica, sendo $Q=-\frac{a^{\frac{3\omega-1}{2}}}{4R}\frac{\partial}{\partial a}\left[a^{\frac{3\omega-1}{2}}\frac{\partial R}{\partial a}\right]$ um potencial de caráter quântico. Tal potencial é o responsável por alterar a trajetória do fator de escala $a$, substituindo a singularidade inicial do Universo por um fator de escala finito. Por sua vez, a equação \eqref{eqbohm2} toma a forma de uma equação de continuidade para $\rho$.

A equação \eqref{WDWcos0} pode ser escrita de forma mais simples considerando a seguinte transformação de coordenadas
\begin{equation}\label{transf}
    \chi=\frac{2}{3(1-\omega)}a^{\frac{3(1-\omega)}{2}},
\end{equation}
a qual resulta em
\begin{equation}\label{WDWcos}
    i \frac{\partial\Psi}{\partial T}=\frac{1}{4}\frac{\partial^2\Psi}{\partial\chi^2}.
\end{equation}
A mesma tranformação modifica \eqref{eqguia} para
\begin{equation}\label{eqguia2}
    \frac{d\chi}{dT}=-\frac{1}{2}\frac{\partial S}{\partial \chi}.
\end{equation}
No contexto da Cosmologia Quântica, $\Psi$ é chamada de função de onda do Universo, a qual deve satisfazer \eqref{WDWcos}. A fim de resolver essa equação para $\Psi$, necessitamos de uma condição de contorno. Uma escolha interessante é dada por
\begin{equation}\label{ccontorno}
\left(\Psi^{*}\frac{\partial\Psi}{\partial\chi}-\Psi\frac{\partial\Psi^{*}}{\partial\chi}\right)\biggl|_{\chi=0}\biggr.=0,
\end{equation}
onde $\Psi^{*}$ representa o conjugado de $\Psi$, uma vez que ela leva a soluções unitárias da função de onda\footnote{Chamamos de unitárias as funções de onda cuja evolução temporal é representada por um operador unitário, o que está intimamente ligado às probabilidades de medidas.}. 

Para a função de onda inicial podemos escolher
\begin{equation}
    \Psi_0=\left(\frac{8}{\pi \sigma^2}\right)^{\frac{1}{4}} \exp\left(-\frac{\chi^2}{\sigma^2}\right),
\end{equation}
uma vez que a condição de contorno \eqref{ccontorno} é satisfeita. A expressão da função de onda para qualquer tempo $T$ é então dada por
\begin{eqnarray}\label{psiT}
    \Psi(\chi,T)=\int_0^\infty G(\chi,\chi_0,T)\Psi_0(\chi_0,T)d\chi_0,
\end{eqnarray}
sendo $G(\chi,\chi_0,T)$ o propagador referente à equação de Wheeler-DeWitt \eqref{WDWcos}. Note que a última é similar à equação de Schr{\"o}dinger, a menos do sinal da energia cinética. Dessa forma, o propagador é dado por
\begin{eqnarray}\label{prop}
 \nonumber   G(\chi,\chi_0,T)&=&\sqrt{-\frac{i}{\pi T}}\left\{\exp\left[-i\frac{(\chi-\chi_0)^2}{T}\right]\right.+\\
    &+&\left.\exp\left[-i\frac{(\chi+\chi_0)^2}{T}\right]\right\},
\end{eqnarray}
onde, a fim de garantir a unitariedade da evolução, somamos um propagador para $\chi_0$ e outro para $-\chi_0$. Ao aplicar \eqref{prop} em \eqref{psiT}, obtemos
\begin{eqnarray}\label{wavef}
 && \nonumber  \Psi(\chi,T)=\left[\frac{8\sigma^2}{\pi(\sigma^4+T^2)}\right]^{\frac{1}{4}}\exp\left[-\frac{\sigma^2\chi^2}{\sigma^4+T^2}\right]\times\\
    &\times& \left[-i\left(\frac{T\chi^2}{\sigma^4+T^2}+\frac{1}{2}\arctan{\frac{\sigma^2}{T}}-\frac{\pi}{4}\right)\right],
\end{eqnarray}
a qual pode ser decomposta no formato $\Psi=R \exp{[iS]}$.

\begin{figure}[h]
	\centering
	\includegraphics[width=0.8\linewidth]{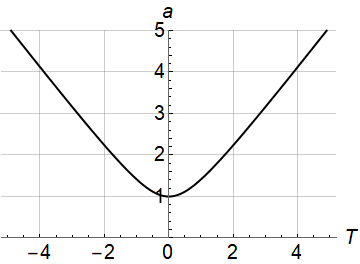}
	\caption{Resolução da singularidade inicial através de um modelo de ricochete obtido na Cosmologia Quântica de De Broglie-Bohm. Nesta figura foram considerados $a_b=\sigma=1$ e $\omega=1/3$, representando um fluido perfeito de radiação.}
	\label{fig:dbb}
\end{figure}

Resolvendo a equação guia \eqref{eqguia2}, obtemos
\begin{eqnarray}
    \chi=\chi_b\left[1+\left(\frac{T}{\sigma^2}\right)^2\right]^{\frac{1}{2}},
\end{eqnarray}
a qual está relacionada ao fator de escala $a$ pela transformação \eqref{transf}, resultando em
\begin{eqnarray}\label{fesc}
    a=a_b\left[1+\left(\frac{T}{\sigma^2}\right)^2\right]^{\frac{1}{3(1-\omega)}}.
\end{eqnarray}
Os parâmetros $\chi_b$ e $a_b$ representam os valores de $\chi$ e $a$ quando o Universo tem seu menor tamanho. A Figura \ref{fig:dbb} apresenta o fator de escala \eqref{fesc} como uma função do tempo $T$, onde podemos identificar um regime de contração do Universo para $T<0$ e um regime de expansão para $T>0$. O encontro dessas fases em $T=0$ se dá no âmbito quântico, onde o potencial $Q$ identificado em \eqref{eqbohm1} desempenha um papel fundamental. Dessa forma, a singularidade clássica é substituída por um fator de escala mínimo $a_b$ (relacionado ao tamanho mínimo do Universo), o que é conhecido na literatura como modelo de ricochete ou {\it bounce}.  

\subsection{Aspectos introdutórios da Cosmologia Quântica em Laços}
A Cosmologia Quântica em Laços se baseia em procedimentos da Gravitação Quântica em Laços adaptados para o caso da métrica de FLRW. Nessa seção utilizaremos a usual interpretação de Copenhagen da Mecânica Quântica, a qual leva a alguns problemas conceituais que serão discutidos na Seção \ref{probmedida}.

Os vínculos de Gauss e do supermomento em \eqref{vincLQG} são trivialmente satisfeitos, enquanto a super-hamiltoniana pode ser escrita como
\begin{eqnarray}\label{shlqc}
    \mathcal{H}=-\frac{1}{\gamma^2} \epsilon_{\bar{i}\bar{j}\bar{k}}\frac{F^{\bar{i}}_{ij}E^{i\bar{j}}E^{j\bar{k}}}{\sqrt{\bar{E}}}.
\end{eqnarray}
A hamiltoniana do sistema pode então ser obtida através da integração de \eqref{shlqc}, a qual é realizada em uma região finita do espaço-tempo a fim de evitar divergências. 

Por sua vez, as variáveis $A^{\bar{i}}_{i}$ e $E^{i}_{\bar{i}}$ podem ser escritas em termos de novas variáveis $c$ e $p$ e do volume $V_0$ dessa região com respeito a uma métrica fiducial $h^{(0)}_{ij}$ definida a partir da co-tríade fiducial $e^{(0) \bar{i}}_i\equiv \delta^{\bar{i}}_i$ como
\begin{equation}
    h^{(0)}_{ij}\equiv e^{(0) \bar{i}}_i e^{(0) \bar{j}}_j \delta_{\bar{i}\bar{j}}.
\end{equation}
As variáveis canônicas tomam então a seguinte forma:
\begin{eqnarray}
    A^{\bar{i}}_i&=&\frac{c}{V_0^{\frac{1}{3}}} e^{(0)\bar{i}}_i,\\
    E^i_{\bar{i}}&=&\frac{p}{V_0^{\frac{2}{3}}}\sqrt{h^{(0)}} e^{(0) i}_{\bar{i}},
\end{eqnarray}
sendo a variável $p$ relacionada ao fator de escala através de $a=\sqrt{|p|}/V_0^{\frac{1}{3}}$. 

A fim de realizar a quantização através das variáveis independentes da métrica, definimos as holonomias de $c$ e o fluxo relacionado a $p$. Estes podem ser canonicamente transformados para novas variáveis $b$ e $v$ ao considerarmos um {\it loop} quadrado fechado, cuja área mínima está relacionada a um autovalor $\Delta$ advindo do espectro discreto da geometria. Definindo tal área mínima como $\bar{\mu}^2 |p|=\Delta$, o que corresponde à chamada dinâmica melhorada \cite{Ashtekar:2006wn}, é possível escrever as novas variáveis como
\begin{eqnarray}
    b&=&\hbar\frac{\bar{\mu} c}{2},\\
    v&=&\frac{sgn(p)|p|^{\frac{3}{2}}}{2\pi l_P^2 \gamma \sqrt{\Delta}},
\end{eqnarray}
onde $sgn(p)$ se refere ao sinal de $p$, $\hbar$ é a constante de Planck reduzida e $l_P$ o comprimento de Planck.

Como feito na Seção \ref{cosmdbb}, consideramos um campo escalar $\phi$ de matéria para realizar o papel do tempo. Dessa forma, o vínculo da super-hamiltoniana \eqref{shlqc} recebe uma contribuição extra dada por \eqref{HTOTAL} com $n=\omega=1$. Por fim, a hamiltoniana H constituída pela contribuição gravitacional e pela contribuição da matéria pode ser promovida a um operador, como mostrado de forma detalhada em \cite{Banerjee:2011qu}. 

\begin{figure}[h]
	\centering
	\includegraphics[width=0.8\linewidth]{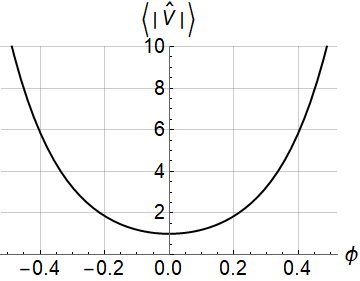}
	\caption{Resolução da singularidade inicial através de um modelo de ricochete obtido na Cosmologia Quântica em Laços. Nesta figura foi considerado $V_b=l_P=1$.}
	\label{fig:lqc}
\end{figure}

Uma maneira de visualizar a resolução da singularidade inicial \cite{Banerjee:2011qu, Ashtekar:2007em} é considerar a função lapso $N$ como o volume $V\equiv|p|^{\frac{3}{2}}$ da região a ser quantizada, o qual é proporcional ao valor absoluto de $v$. Dessa forma, podemos promover $b$ e $v$ a operadores $\hat{b}$ e $\hat{V}=-i\hbar \partial_b$ e escolher um ordenamento de forma que o vínculo da super-hamiltoniana assuma a forma
\begin{equation}
    \left[3\pi l_P^2 \sin^2{(2b\partial_b)}-\partial^2_\phi\right]\Psi=0.
\end{equation} 
Nesse caso, é possível mostrar \cite{Ashtekar:2007em} que o valor esperado, na interpretação de Copenhagen, do volume $\hat{V}$ é dado por
\begin{equation}
     \left< |\hat{V}| \right>=V_b \cosh{\left(\sqrt{12\pi l_P^2}\phi\right)},
\end{equation}
sendo $V_b$ o volume mínimo do Universo. A Figura \ref{fig:lqc} mostra o comportamento de $\left< |\hat{V}| \right>$ em função do tempo $\phi$.

\subsection{O problema do tempo}\label{ProbOfTime}
Um dos problemas conceituais da Gravitação e da Cosmologia quânticas é o chamado problema do tempo, o qual está relacionado ao caráter covariante da Teoria da Relatividade Geral. Na Mecânica Quântica usual, o tempo é um parâmetro externo, distinto das componentes do sistema que são quantizadas. Dessa forma, a maneira através da qual o tempo é representado e entendido em uma teoria da Gravitação Quântica ainda não é bem estabelecida. 

Matematicamente, podemos ter uma ideia desse problema através da equação de Wheeler-DeWitt \eqref{Ceq1}. Tendo em vista sua relação com a covariância da teoria frente a transformações da coordenada temporal, é esperado que essa equação forneça a dinâmica do funcional de onda $\Psi$ no tempo. Na equação de Schr{\"o}dinger, o tempo aparece como uma primeira derivada da função de onda e, portanto, buscamos por um termo similar em \eqref{Ceq1}. Entretanto, no geral esse termo não aparece diretamente. Além disso, tendo em vista que a hamiltoniana do sistema é constituída pelos vínculos, ao aplicá-los em \eqref{Seq} obtemos $i\partial_0 \Psi=H=0$, indicando que o funcional de onda $\Psi$ não depende do tempo.

No contexto da Cosmologia Quântica, uma possível solução para esse problema é atribuir o tempo a graus de liberdade relacionados à matéria que preenche o Universo, de forma a obter um termo proporcional ao momento conjugado. Isso pode ser visto diretamente na equação \eqref{HTOTAL}, onde o momento conjugado $P_T$, relacionado ao campo escalar $\phi$ que descreve um fluido perfeito, aparece linearmente. 

Assim como as abordagens para quantização, diversas soluções para o problema do tempo são investigadas. Alguns exemplos podem ser encontrados em \cite{Christodoulakis_2011,Isham1993,Anderson:2012vk}.

\subsection{O problema da medida}\label{probmedida}
Outro problema conceitual de extrema importância para a Cosmologia Quântica é o problema da medida, o qual surge na interpretação de Copenhagen como resultado do colapso da função de onda. Mais especificamente, a interpretação de Copenhagen descreve o sistema quântico através de uma função de onda contendo uma superposição linear dos estados possíveis para o sistema. Quando uma medida é realizada por um observador, a função de onda colapsa em um único estado, o qual corresponde à realidade percebida pelo observador. Entretanto, diferentemente da dinâmica da função de onda antes da medida, o colapso não é descrito pela equação de Schrödinger. Em outras palavras, o colapso da função de onda nada mais é que um postulado. Tal descrição implica no surgimento de sérios questionamentos sobre a natureza ontológica da Mecânica Quântica. Afinal de contas, o postulado da medida implicaria na inexistência de uma realidade objetiva independente de observações. Como definir então o que configura uma medida e quais entidades possuem o status de observador? No contexto da Cosmologia Quântica tal problema é ainda mais agravado, uma vez que a existência do Universo em um determinado estado quântico dependeria de um observador externo, o qual não pode existir por definição. 

A natureza determinística da interpretação de De Broglie-Bohm leva, automaticamente, à solução do problema da medida. Nesta interpretação, a função de onda é constituída por ramos incomunicáveis, sendo somente um deles selecionado pelas condições iniciais do sistema. Dessa forma, toda a dinâmica do sistema é descrita pela Mecânica Bohmiana, sem a necessidade de postular o colapso da função de onda. Mais detalhes sobre o processo de medida na interpretação de De Broglie-Bohm podem ser encontrados em \cite{holland1995quantum}.

Uma outra proposta para resolver o problema da medida é a chamada decoerência, a qual faz uso da interação do sistema com o ambiente a fim de selecionar um estado da função de onda \cite{SCHLOSSHAUER20191}. Tal descrição fornece uma explicação para a transição do âmbito quântico para o âmbito clássico, mas não aborda a unicidade dos acontecimentos, uma vez que todos os ramos da função de onda existem simultaneamente.

\section{Conexões com observáveis}\label{VinculosObs}
Tendo em vista a escala de energia em que efeitos quânticos devem se tornar relevantes e a nossa atual limitação em obter informações de épocas anteriores à CMB, a conexão das teorias de Cosmologia Quântica com observáveis se mostra desafiadora. Apesar disso, algumas propostas para testar tais modelos já existem na literatura, inclusive utilizando dados já coletados. Nesta seção serão apresentadas algumas dessas propostas.

Uma vez que os fótons da CMB são atualmente a mais antiga fonte de informação à qual temos acesso, é natural buscar por evidências observacionais da Cosmologia Quântica nessa radiação de fundo. Devido à esfericidade do céu que observamos, as propriedades da CMB são usualmente descritas através de uma decomposição esférica. Dessa forma é possível quantificar correlações entre diferentes pontos no céu e relacioná-las a perturbações cosmológicas\footnote{As perturbações cosmológicas são pequenos desvios em torno da métrica de fundo do espaço-tempo, podendo ser separadas em escalares, vetoriais e tensoriais (ondas gravitacionais primordiais). Tais perturbações são responsáveis por gerar as estruturas que observamos hoje e estão diretamente relacionadas às flutuações observadas na CMB.} geradas por modelos de Universo primordial, incluindo modelos de Cosmologia Quântica. Os chamados espectros de potência representam quantidades de extrema relevância nesse contexto. Através deles acessamos as amplitudes e os chamados índices espectrais\footnote{Os espectra de potência escalar e tensorial são recpectivamente parametrizados por $P_s=A_s\left(
k/k_*\right)^{n_s-1}$ e $P_t=A_t\left(
k/k_*\right)^{n_t}$, sendo $A_s$ e $A_t$ suas amplitudeds e $n_s$ e $n_t$ seus índices espectrais.} de perturbações escalares e tensoriais. A amplitude das perturbações tensoriais possui um limite superior advindo do fato de que estas ainda não foram observadas, enquanto o espectro de potência escalar que observamos é quase invariante de escala ($n_s-1\approx 0$) e inclinado para o vermelho ($n_s-1< 0$). Tais fatos podem ser usados para restringir modelos de Cosmologia Quântica, como mostrado em \cite{Cai:2014bea,Cai:2014xxa,Battefeld:2014uga, Calcagni:2012vw}. 

Além dos espectros de potência, os biespectros também representam uma maneira eficiente de restringir os modelos, sendo relacionados a correlações entre três diferentes pontos no céu. Tais correlações fornecem informação sobre as chamadas não-gaussianidades, cujas características dependem do modelo de Universo primordial considerado. Uma interessante conexão com observáveis foi obtida por modelos de ricochete motivados pela Cosmologia Quântica em Laços. Foi proposto que tais cenários com grandes não-gaussianidades seriam capazes de aliviar as chamadas anomalias em grandes escalas da CMB\footnote{As anomalias em grandes escalas são características da CMB que observamos e que possuem uma pequena probabilidade de acontecer no Modelo Cosmológico Padrão.} \cite{Agullo_2021}. Entretanto, os limites impostos por dados do Planck sobre o biespectro de temperatura excluem tal possibilidade \cite{Delgado:2021mxu, vanTent:2022vgy}. Ainda no contexto da Cosmologia Quântica em Laços, um template oscilatório para o biespectro primordial foi investigado em \cite{K:2023gsi}, juntamente com o biespectro da CMB correspondente. 

Além dos cenários de ricochete, outros aspectos da Cosmologia Quântica podem ser investigados através de observáveis. Um importante exemplo é dado pela possibilidade de testar a decoerência através dos espectros da CMB. Tendo em vista que as perturbações cosmológicas possuem uma origem quântica, elas estariam sujeitas ao fenômeno de ``classicalização'' descrito pela decoerência. Como consequência, impressões desse processo são esperadas na CMB, como discutido em \cite{Martin:2018zbe,DaddiHammou:2022itk}. 

Assim como os fótons da CMB, ondas gravitacionais primordiais podem formar uma radiação (gravitacional) de fundo, a qual tem sido procurada por diferentes {\it surveys}, tais como os {\it Pulsar Timing Arrays} (PTAs) e os futuros interferômetros espaciais. Tais ondas gravitacionais estão acopladas à matéria e à radiação de forma fraca, o que as possibilita viajar livremente desde sua formação. Dessa forma, ondas gravitacionais primordiais podem se propagar até nós desde os primórdios do Universo, inclusive antes da CMB. Se detectado, tal fundo estocástico de ondas gravitacionais poderia ser utilizado para testar diferentes cenários de Cosmologia Quântica, como discutido em \cite{Calcagni_2021}. 

Dessa forma, apesar dos desafios impostos pelas escalas de energia associadas a cenários de Cosmologia Quântica, diferentes abordagens têm contribuído para o progresso da fenomenologia da área. Além das restrições já existentes graças à detecção da CMB, futuros limites são esperados da detecção do fundo estocástico de ondas gravitacionais, o que representaria uma revolução para a física do Universo primordial. 

\section*{Agradecimentos}
Gostaria de agradecer à Carla R. Almeida, ao Lucca Fazza e ao Nelson Pinto-Neto pelas discussões e ao financiamento do {\it National Science Centre}, Polônia, projeto No. UMO-2018/30/Q/ST9/00795.\\\\

$^\dagger$\href{mailto:paola.moreira.delgado@doctoral.uj.edu.pl}{paola.moreira.delgado@doctoral.uj.edu.pl}\\

\newpage

\bibliography{main} 

\end{document}